# Explicit drain-current model of graphene field-effect transistors targeting analog and radio-frequency applications

David Jiménez and Oana Moldovan

Departament d'Enginyeria Electrònica, Escola d'Enginyeria,
Universitat Autònoma de Barcelona, 08193-Bellaterra, Spain

david.jimenez@uab.es

**Abstract**

We present a compact physics-based model of the current-voltage characteristics of graphene field-effect transistors, of especial interest for analog and radio-frequency applications where bandgap engineering of graphene could be not needed. The physical framework is a field-effect model and drift-diffusion carrier transport. Explicit closed-form expressions have been derived for the drain current covering continuosly all operation regions. The model has been benchmarked with measured prototype devices, demonstrating accuracy and predictive behavior. Finally, we show an example of projection of the intrinsic gain as a figure of merit commonly used in RF /analog applications.

Graphene has emerged as a candidate material for future nanoelectronic devices. It exhibits very high mobility ($\sim 10^5$ cm$^2$/V-s) and saturation velocity ($\sim 10^8$ cm/s), together with a promising ability to scale to short gate lengths and high speeds by virtue of its thinness. The main concern is the absence of a gap, therefore limiting the usefulness in digital applications. Several approaches have appeared in the last years to open a gap. For instance, graphene nanoribbons, bilayer graphene, or strained graphene have been suggested. However, graphene could still be very useful in analog and radiofrequency (RF) applications where high ON/OFF current ratios are not required [1]. In small signal amplifiers, for instance, the transistor is operated in the ON-state and small RF signals that are to be amplified are superimposed onto the DC gate-source voltage. Instead, what is needed to push the limits of many analog/RF figures of merit, for instance the cut-off frequency or the intrinsic gain, is an operation region where high transconductance together with a small output conductance is accomplished. These conditions are realized for state-of-the-art graphene field-effect transistors (GFETs). Specifically, for large-area GFETs, the output characteristic shows a weak saturation that could be exploited for analog/RF applications. Using this technology, cut-off frequencies in the THz range are envisioned. Actually, cut-off frequencies in the range of hundreds of GHz have been demonstrated using non-optimized technologies [2,3].

At this stage of development of GFET technology, modeling of the current-voltage (I-V) characteristics is important for device design optimization, projection of performances, and exploration of analog / RF circuits providing new or improved functionalities. Here we present a compact physics-based model of the I-V characteristics of GFETs where explicit closed-form expressions have been derived for the drain current. The framework is a field-effect model and drift-diffusion carrier transport, which is accurate to explain the electrical behavior of GFETs [4,5]. The compact model has been benchmarked with measured I-V characteristics of prototype devices, demonstrating accuracy and predictive behavior.

In the following we focus on GFET with the cross-section depicted in Fig. 1. The electrostatics of this device can be understood using the equivalent capacitive circuit from [5]. There, $C_t$ and $C_b$ are the top and bottom oxide capacitances and $C_q$ represents the quantum capacitance of the graphene. The potential $V_c$ represents the voltage drop across $C_q$, where $C_q = k|V_c|$ under the condition $qV_c \gg k_B T$, with $k = \frac{2q^2}{\pi} \frac{q}{(\hbar v_F)^2}$, and $v_F$ (=$10^6$ m/s) is the Fermi velocity [6]. The potential $V(x)$ is the voltage drop in the graphene channel, which is zero at the source end at x=0 and equal to the drain-source voltage ($V_{ds}$) at the drain end at x=L. Applying circuit laws to the equivalent capacitive circuit and noting that the overall net mobile sheet charge density in the graphene channel, defined as $Q_c$=q(p-n), is equal to -(1/2)$C_q V_c$, the following relation is obtained:

$$V_c(x) = \left(V_{gs} - V_{gs0} - V(x)\right)\frac{C_t}{C_t + C_b + \frac{1}{2}C_q} + \left(V_{bs} - V_{bs0} - V(x)\right)\frac{C_b}{C_t + C_b + \frac{1}{2}C_q} \quad (1)$$

where $V_{gs}$-$V_{gs0}$ and $V_{bs}$-$V_{bs0}$ are the top and back gate-source voltage overdrive, respectively. These quantities comprise work-function differences between the gates and the graphene channel, eventual charged interface states at the graphene/oxide interfaces, and possible doping of the graphene.

To model the drain-current a drift-diffusion carrier transport is assumed under the form $I_{ds} = -qW\rho_c(x)v(x)$, where W is the gate width, $\rho_c(x)$=|$Q_c$|/q is the free carrier sheet density in the channel at position x, and v(x) the carrier drift velocity. Using a soft-

saturation model, consistent with Monte Carlo simulations [7], v(x) can be expressed as v=μE/(1+μ|E|/$v_{sat}$), where E is the electric field, μ is the carrier low-field mobility, and $v_{sat}$ is the saturation velocity. The latter is concentration-dependent and given by $v_{sat} = \Omega/\sqrt{\pi \rho_c}$ [4]. Applying E=-dV(x)/dx, combining the above expressions for v and $v_{sat}$, and integrating the resulting equation over the device length, the drain current becomes:

$$I_{ds} = \frac{q\mu W \int_0^{V_{ds}} \rho_c dV}{L + \mu \left| \int_0^{V_{ds}} \frac{1}{v_{sat}} dV \right|} \quad (2)$$

The denominator represents an effective length ($L_{eff}$) to take into account the saturation velocity effect. In order to get an explicit expression for $I_{ds}$, the integrals in (2) are solved using $V_c$ as the integration variable and consistently expressing $\rho_c$ and $v_{sat}$ as a function of $V_c$:

$$I_{ds} = \frac{q\mu W \int_{V_{cs}}^{V_{cd}} \rho_c(V_c) \frac{dV}{dV_c} dV_c}{L + \mu \left| \int_{V_{cs}}^{V_{cd}} \frac{1}{v_{sat}(V_c)} \frac{dV}{dV_c} dV_c \right|} \quad (3),$$

where $V_c$ is obtained from (1) and can be written as:

$$V_c = \frac{-(C_t + C_b) + \sqrt{(C_t + C_b)^2 \pm 2k[(V_{gs} - V_{gs0} - V)C_t + (V_{bs} - V_{bs0} - V)C_b]}}{\pm k} \quad (4).$$

The positive (negative) sign applies whenever $(V_{gs} - V_{gs0} - V)C_t + (V_{bs} - V_{bs0} - V)C_b > 0\ (< 0)$. The channel potential at the source ($V_{cs}$) is determined as $V_c(V=0)$. Similarly, the channel potential at the drain ($V_{cd}$) is determined as $V_c(V=V_{ds})$. Moreover, (1) provides the relation $\frac{dV}{dV_c} = -\left(1 + \frac{kV_c sgn(V_c)}{C_t + C_b}\right)$, where *sgn* refers to the sign function. On the other hand, the charge sheet density can be written as $\rho_c(V_c) = kV_c^2/(2q) + \rho_0$. The extra term $\rho_0$ added to $\rho_c$ accounts for the carrier density induced by impurities [8]. Inserting these expressions into (3), the following explicit drain current expression can be finally obtained:

$$I_{ds} = \mu \frac{W}{L_{eff}} \left\{ q\rho_0 V_{ds} - \frac{k}{6}(V_{cd}^3 - V_{cs}^3) - \frac{k^2}{8(C_t + C_b)} \left( sgn(V_{cd})V_{cd}^4 - sgn(V_{cs})V_{cs}^4 \right) \right\}$$

$$L_{eff} = L + \mu \frac{\sqrt{\pi}}{\Omega} \left| \left\{ \begin{array}{c} -\frac{V_c}{2}\sqrt{\frac{kV_c^2}{2q} + \rho_0} - \frac{\rho_0 \log\left(2\left(\sqrt{\frac{k}{2q}}\sqrt{\frac{kV_c^2}{2q} + \rho_0} + \frac{kV_c}{2q}\right)\right)}{2\sqrt{\frac{k}{2q}}} \\ -sgn(V_c)\frac{2q}{3(C_t + C_b)}\left(\left(\frac{kV_c^2}{2q} + \rho_0\right)^{\frac{3}{2}} - \rho_0^{\frac{3}{2}}\right) \end{array} \right\} \right|_{V_{cs}}^{V_{cd}} \quad (5)$$

To reproduce the experimental I-V characteristics, accounting for the voltage drop at the S/D contacts is necessary. This quantity must be removed from the external $V_{ds\_ext}$ in order to get the internal $V_{ds}$. This is done by solving the equation $V_{ds} = V_{ds\_ext} - I_{ds}(V_{ds})(R_s + R_d)$, noting that $I_{ds}$ is a function of $V_{ds}$ as given by (5).

To test the model we have benchmarked the resulting I-V characteristics with experimental results extracted from devices in [4,9]. The first device under test has *L*=1 μm, W=2.1 μm, top dielectric is HfO$_2$ of 15 nm and permittivity ~ 16, and the bottom dielectric is silicon oxide of 285 nm [4]. The backgate voltage was -40 V. The flat-band voltages $V_{gs0}$ and $V_{bs0}$ were tuned to 1.45 V and 2.7 V, respectively. These values were selected to locate the Dirac point according to the experiment. It is worth noting that a ratio $(C_t//C_q)/C_b \sim 46$ was estimated measuring the top-gate Dirac point at different back-gate voltage. To capture this ratio, responsible of the gate efficiency, an effective top dielectric thickness of 26 nm was used, yielding a better fit with the experiments. A low-field mobility of 1200 cm$^2$/V-s for both electrons and holes, S/D resistance of 800 $\Omega$, and phonon effective energy $\hbar\Omega$~55 meV were considered. These values are consistent with the extracted values. A sheet carrier density $\rho_0$=10$^{11}$ cm$^{-2}$ was selected for the final fine tuning.

Now we consider the output characteristics (Fig. 2). To fit the experiment, a phonon energy $\hbar\Omega$~15 meV was used for V$_{gs}$-V$_{gs0}$<<0. This is justified because the phonon effective energy overestimates the actual value for high sheet carrier densities [10]. Due to the gapless channel the output characteristics present a saturation-like behavior that includes a second linear region. The cross-over between the first and second linear region could be understood observing V$_c$ as a function of V$_{ds\_ext}$ (see the inset). Here V$_{cd}$ and V$_{cs}$ are both negative for small values of V$_{ds\_ext}$, meaning that the channel is entirely p-type. Increasing V$_{ds\_ext}$, at some point V$_{cd}$ becomes positive and the channel switches to n-type near the drain end.

The second examined device is a top-gate device using a different fabrication technology with L=10 $\mu$m, W=5 $\mu$m, and HfO$_2$ as a dielectric with thickness of 40 nm [9]. The flat-band voltage was V$_{gs0}$=0.85 V according to the experiment. A low-field mobility of 7500 cm$^2$/V-s for both electrons and holes, S/D resistance of 300 $\Omega$, and $\hbar\Omega$~100 meV were considered. A sheet carrier density $\rho_0$=3·10$^{11}$ cm$^{-2}$ was used for the final tuning.

The resulting I-V characteristics are shown in Fig. 3. We have extended the simulated voltage range beyond the experiment range to show the predictive behavior of the model. The transfer characteristics exhibit an ambipolar behavior dominated by holes (electrons) for V$_{gs}$-V$_{gs0}$<V$_{gs,D}$ (>V$_{gs,D}$) where V$_{gs,D}$ (Dirac gate voltage) is given by V$_{gs,D}$=V$_{gs0}$+V$_{ds}$/2. The output characteristics behave similar as the first examined device. Once again, the comparison between the model and the experiment further demonstrates the model accuracy.

Next, we will give an example on how to use our current-voltage DC model to project an important figure-of-merit (FoM) used in RF / analog applications, namely the intrinsic gain ($G = g_m/g_{ds}$), which is defined as the ratio of the transconductance ($g_m = \frac{\partial I_{ds}}{\partial V_{gs}}$) and output conductance ($g_{ds} = \frac{\partial I_{ds}}{\partial V_{ds}}$). Both $g_m$ and $g_{ds}$ are small-signal quantities directly derived from the DC model. Fig. 4a shows projection of these quantities together with G, for the transistor topology from Ref. [9], as a function of $V_{ds}$ (internal) for a fixed $V_{gs}$=-0.75 V. Remarkably, when the transistor is operated at the beginning of the saturation region ($V_{ds,p\text{-off-d}}=V_{gs}-V_{gs0}$=-1.6 V), $g_m$ is maximum while $g_{ds}$ is minimum, yielding an optimal G (=15). This particular value $V_{ds,p\text{-off-d}}$ sets the pinch-off point exactly at x=L indicating the cross-over between a p-type channel and a mixed p-type / n-type ambipolar channel. As a peculiar characteristic of GFETs, different from what it is observed in conventional silicon FETs, there exists a region of negative $g_m$ (and G) for $V_{ds} < V_{ds,D} = 2V_{ds,p\text{-off-d}}$ (=-3.2 V), where $V_{ds,D}$ could be named as Dirac drain voltage. At this particular voltage, one half of the channel behaves as n-type and the other as p-type. The negative $g_m$ is a consequence of the crossing I-$V_{ds}$ characteristics. It seems undesirable for normal operation of the transistor to work in this region, although it cannot be ruled out that any practical application could be found. Finally, Fig. 4b shows how the intrinsic gain for the same transistor could be tailored by expanding the range of applied gate and drain voltage. For example, intrinsic gains as large as G ~ 100 could be obtained operating the transistor at $V_{gs}$=-1.75 V and $V_{ds}$ ~ -2.6 V.

In conclusion, we have presented an explicit and compact drain-current model for large-area GFETs based on a field-effect model and drift-diffusion carrier transport, of especial interest as a tool for design of analog / RF applications based on graphene transistor technologies. We have illustrated how to use the current-voltage DC model to find an important FoM used in RF / analog applications; namely the intrinsic gain. Discussions of other FoM like the cutoff frequency can be found in Ref. [5].

**Acknowledgments**

Funding of MICINN under contracts FR2009-0020 and TEC2009-09350, and the DURSI of the Generalitat de Catalunya under contract 2009SGR783 is acknowledged.

**References**

[1] F. Schwierz, "Graphene transistors", Nature Nanotechnology, vol. 5, pp. 487-496 (2010).

[2] Y.-M. Lin, C. Dimitrakopoulos, K. A. Jenkins, D. B. farmer, H.-Y. Chiu, A. Grill, and Ph. Avouris, "100-GHz transistors from wafer-scale epitaxial graphene", Science 327, no. 5966, p. 662 (2010).

[3] L. Liao, Y-C. Lin, M. Bao, R Cheng, J. Bai, Y. Liu, Y. Qu, K. L. Wang, Y. Huan, and X. Duan, "High-speed graphene transistors with a self-aligned nanowire gate ", Nature 467, no. 16, pp. 305-308 (2010).

[4] I. Meric, M. Y. Han, A. F. Young, B. Ozyilmaz, P. Kim, and K. Shepard, "Current saturation in zero-bandgap, top-gated graphene field-effect transistors", Nature Nanotechnology, vol. 3, pp. 654-659, (2008).

[5] S. Thiele, J. A. Schaefer, and F. Schwierz, "Modeling of graphene metal-oxide-semiconductor field-effect transistors with gapless large-area graphene channels", Journal of Applied Physics, vol. 107, 094505 (2010).

[6] T. Fang, A. Konar, H. Xing, and D. Jena, "Carrier statistics and quantum capacitance of graphene sheets and ribbons," Appl. Phys. Lett., vol. 91, 092109 (2007).

[7] J. Chauhan and J. Guo, "High-field transport and velocity saturation in graphene," Appl. Phys. Lett., vol. 95, 023120 (2009).

[8] J. Xia, F. Cheng, J. Li, and N. Tao, "Measurement of the quantum capacitance of graphene," Nature Nanotech., vol. 4, pp. 505-509 (2009).

[9] J. Kedzierski, P-L. Hsu, A. Reina, J. Kong, P. Healey, P. Wyatt, and C. Keast, "Graphene-on-insulator transistors made using C on Ni chemical-vapor deposition," IEEE Electron Device Lett., vol. 30, pp. 745-747, (2009).

[10] I. Meric, C. R. Dean, A. F. Young, J. Hone, P. Kim, and K. Shepard, "Graphene field-effect transistors based on boron nitride gate dielectrics," International Electron Devices Meeting Tech. Digest, pp. 556-559, (2010).

**Figure captions:**

Fig. 1. Cross section of the dual-gate GFET. It consists of a large-area graphene channel on the top of an insulator layer, playing the role of backgate dielectric, grown on a heavily doped Si wafer acting as backgate. The source and drain electrodes are contacting the graphene channel from the top and are assumed to be ohmic. The source is grounded and considered the reference potential in the device. The electrostatic modulation of the carrier concentration in graphene is achieved via a top-gate stack consisting of the gate dielectric and the gate.

Fig. 2. Output characteristics obtained from the analytical model (solid lines) compared with experimental results from Ref. [4] (symbols). Inset: quantum capacitance voltage drop as a function of the external source-drain voltage for different gate voltage overdrive

Fig. 3. Transfer (a) and output (b) characteristics obtained from the compact model (solid lines) compared with experimental results from Ref. [9] (symbols).

Fig. 4. (a) Intrinsic gain, transconductance and output conductance as a function of the drain voltage. The channel is shown to be ambipolar, or p-type, depending on the drain voltage. (b) Intrinsic gain as a function of the drain voltage for different gate voltage to show how this figure of merit can be tailored by properly selecting the bias point.

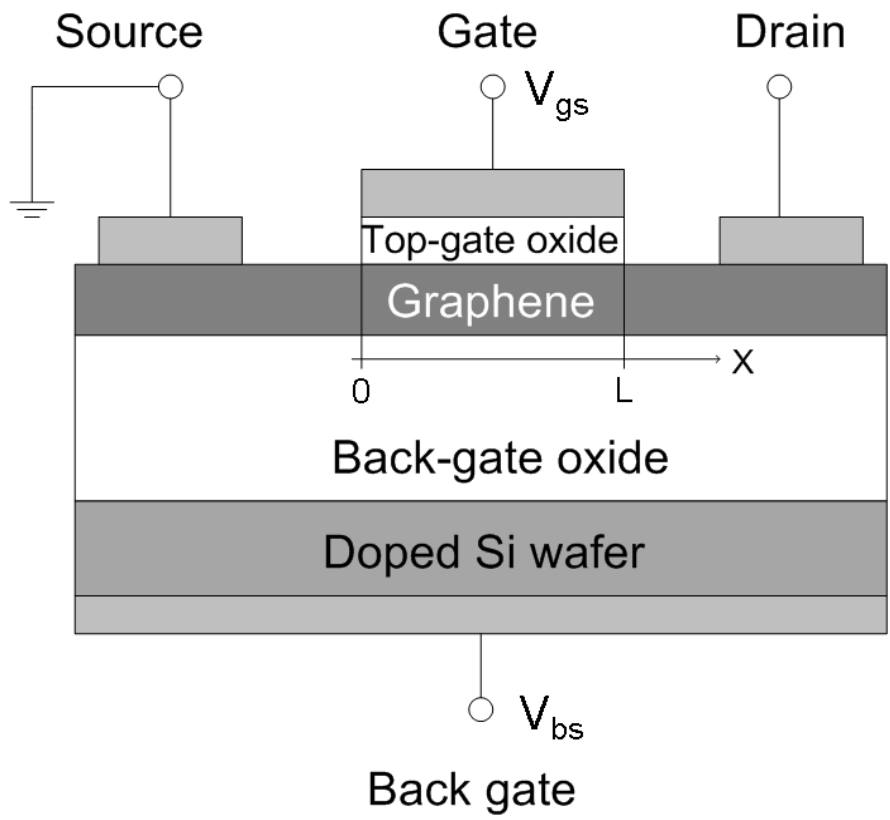

Figure 1

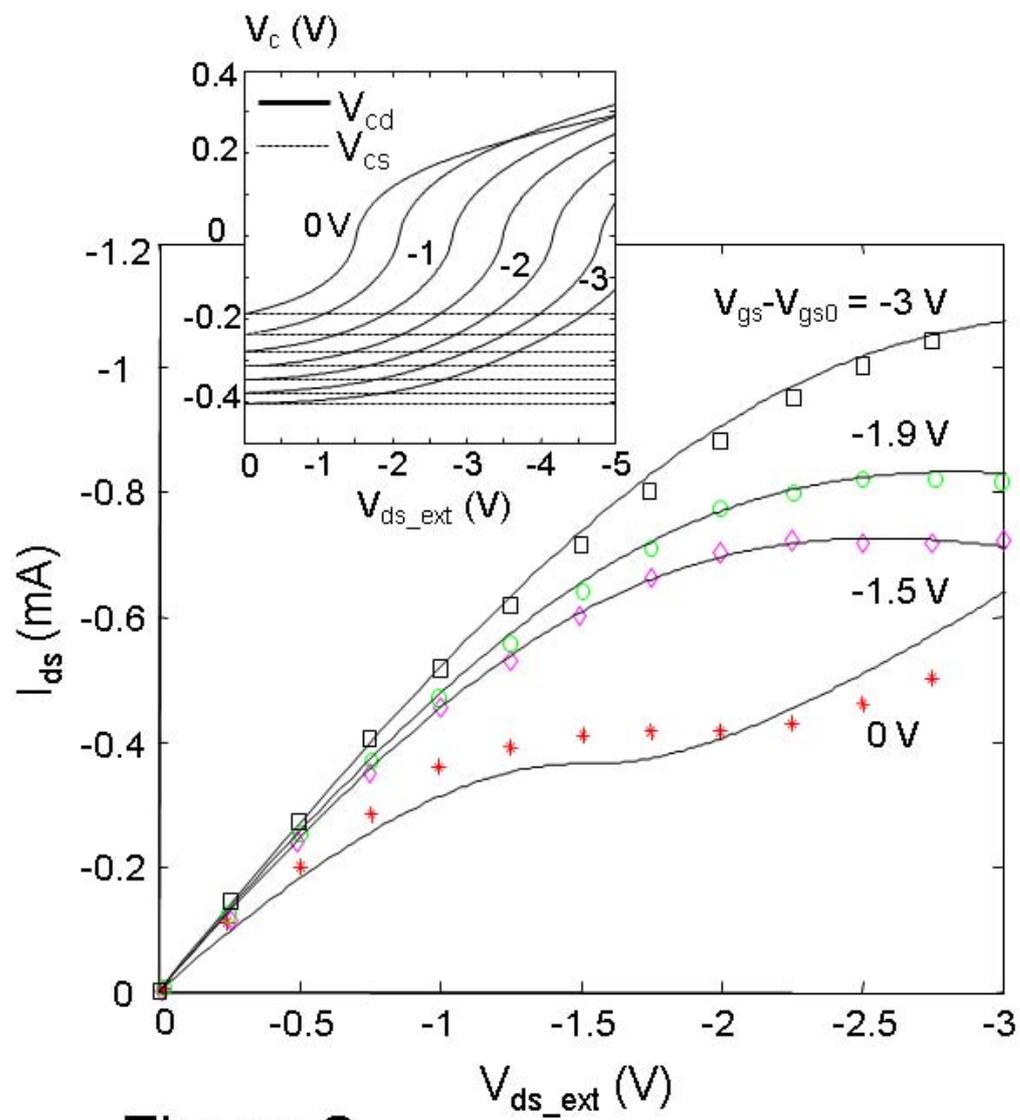

Figure 2

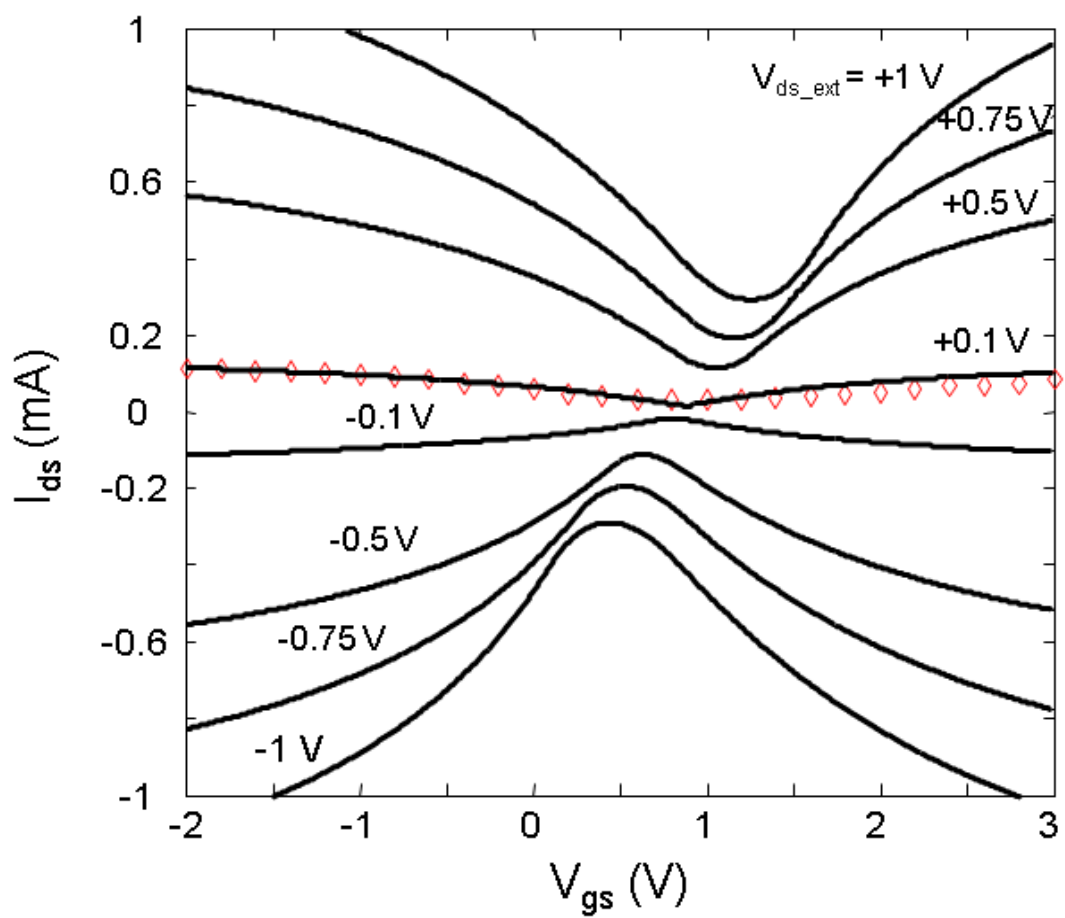

Figure 3a

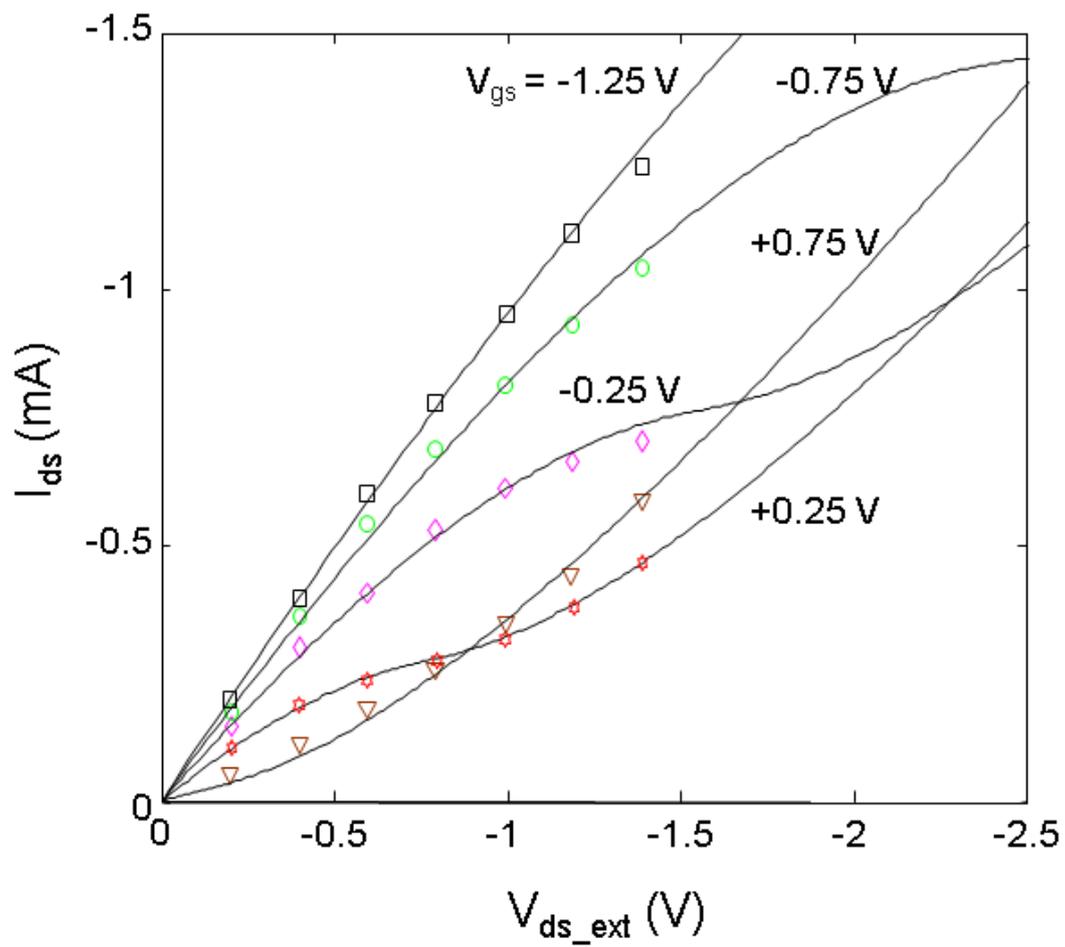

Figure 3b

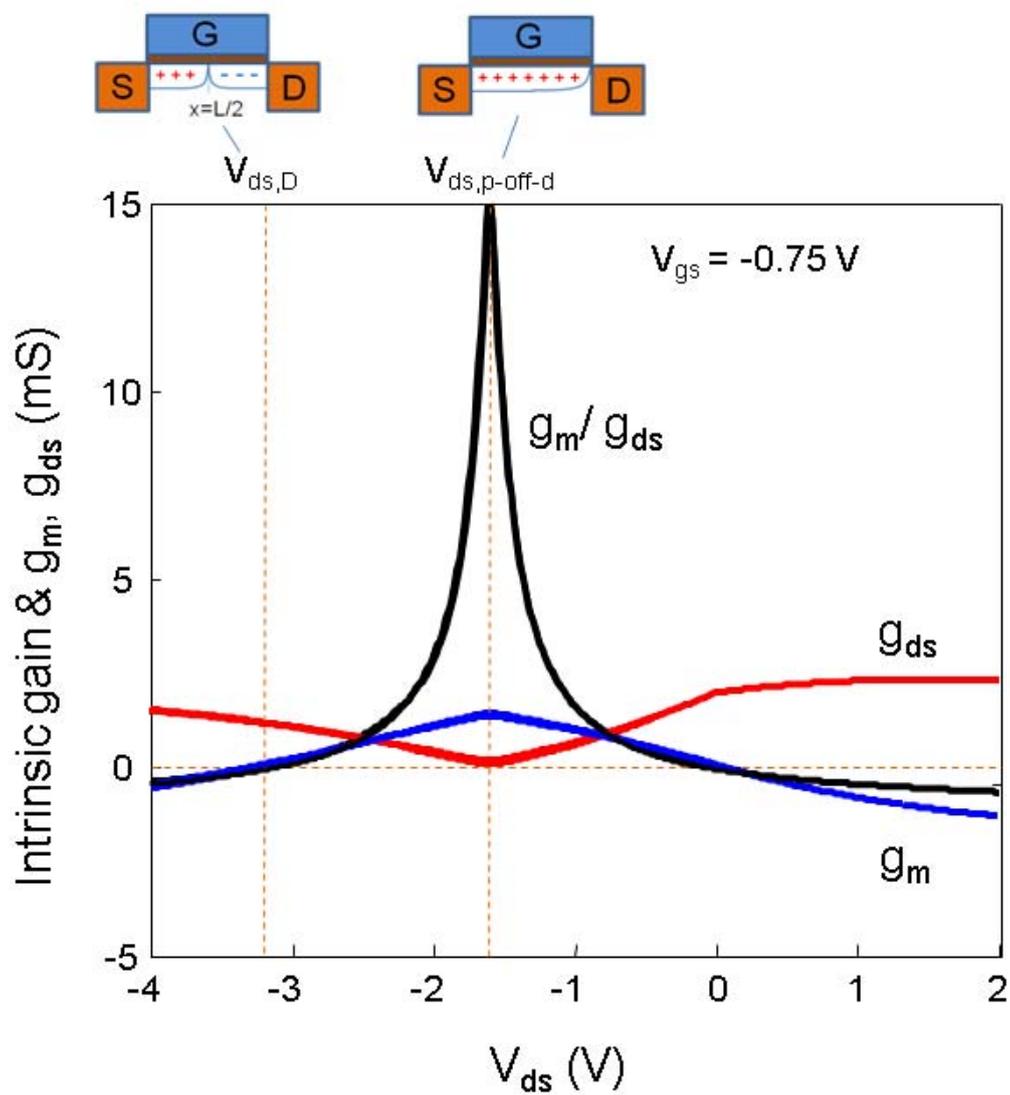

Figure 4a

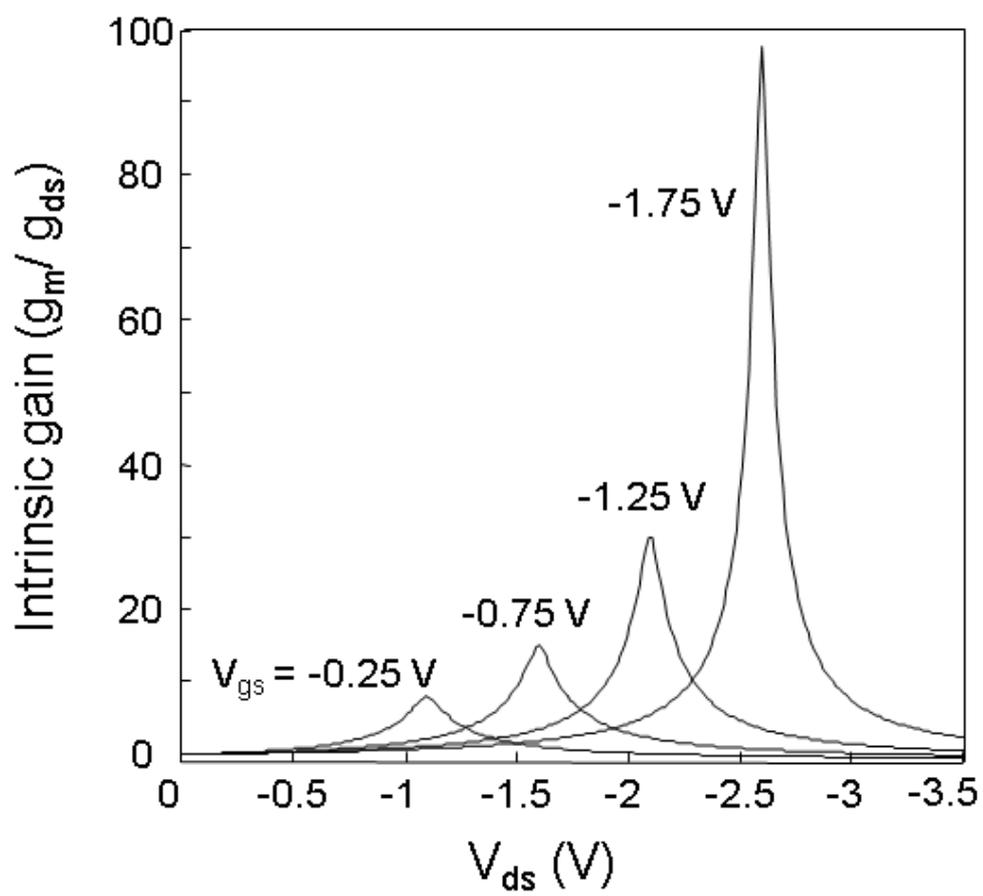

Figure 4b